\documentclass[a4paper,12pt]{article}
\usepackage{amssymb}
\newtheorem{thm}{Theorem}
\newtheorem{prop}{Proposition}
\newtheorem{lemma}{Lemma}
\newtheorem{coro}{Corollary}
\def\ev#1{\overline{#1}}

\def\P{{\mathbb P}}
\def\Z{{\mathbb Z}}
\def\dfrac#1#2{{\displaystyle\frac{#1}{#2}}}
\def\sfrac#1#2{{\frac{#1}{#2}}}
\def\hhf#1#2{F_#1(#2)}
\def\hhg#1#2{G_#1(#2)}
\def\th#1{\vartheta_p(#1)}
\def\midasi#1{\medskip\noindent{\underline{\it #1}.}}
\def\prf{\noindent{\underline{\it Proof}}. }
\def\qed{$\square$}
\begin{document}


\begin{center}
{\Large {\bf Pad\'e interpolation for
elliptic Painlev\'e equation}}\\
\vskip10mm
{\large Masatoshi Noumi$^a$, Satoshi Tsujimoto$^b$ and Yasuhiko Yamada$^a$}\\
\vskip5mm
a) Department of Mathematics, Faculty of Science,\\
Kobe University, Hyogo 657-8501, Japan\\
b) Department of Applied Mathematics and Physics Graduate School of Informatics,\\
Kyoto University, Kyoto 606-8501, Japan\\
\vskip10mm
Dedicated to Professor Michio Jimbo on his 60th birthday
\vskip10mm

\end{center}

\noindent
{\bf Abstract.}
An interpolation problem related to the elliptic Painlev\'e equation 
is formulated and solved.
A simple form of the elliptic Painlev\'e equation and the Lax pair are obtained.
Explicit determinant formulae of special solutions are also given.

\vskip5mm




\section{Introduction}\label{sect:intro}

There exists a close connection between the Painlev\'e equations and the Pad\'e approximations (e.g. \cite{Magnus} \cite{Yfe}).
An interesting feature of the Pad\'e approach to Painlev\'e equation is that we can obtain Painlev\'e equations, its Lax formalism
and special solutions simultaneously once we set up a suitable Pad\'e problem.
This method is applicable also for discrete cases and it gave a hint for a Lax pair \cite{Ysigma} for the elliptic
difference Painlev\'e equation \cite{Sakai}.

In this paper, we analyze the elliptic Painlev\'e equation, its Lax pair and special solutions,
by using the Pad\'e approach. In particular, we study the discrete deformation
along one special direction\footnote{
Though all the directions are equivalent due to the B\"acklund transformations, there exists
one special direction in the formulation on ${\P}^1 \times {\P}^1$ for which the equation take a simple form like QRT system \cite{QRT}.
Jimbo-Sakai's $q$-Painlev\'e six equation \cite{JS} is a typical example of such beautiful equations.
For various $q$-difference cases, the Lax formalisms for such direction were studied in \cite{Yimrn}.}. 
As a result, we obtain remarkably simple form of the elliptic Painlev\'e equation (\ref{eq:fev}), (\ref{eq:gev})
and its Lax pair ({\ref{eq:L1}), (\ref{eq:L2}) or  (\ref{eq:L3}), together with their explicit special solutions
given by equations (\ref{eq:fgbyUV}), (\ref{eq:UVdet}) and (\ref{eq:taushift}).

This paper is organized as follows.
In section \ref{sect:prob}, we set up the interpolation problem.
In section \ref{sect:cont}, we derive two fundamental contiguity relations satisfied
by the interpolating functions.
In section \ref{sect:painl}, we show that the variables $f,g$ appearing in the contiguity relations
satisfy the elliptic Painlev\'e equation.
Interpretation of the contiguity relations as the Lax pair for elliptic Painlev\'e equation
is given in section \ref{sect:lax}.
In section \ref{sect:det}, explicit determinant formulae for the interpolation problem are given.
Derivation of the Painlev\'e equation (\ref{eq:fev}), (\ref{eq:gev}) based on affine Weyl group action
is given in Appendix A.

\section{The interpolation problem}\label{sect:prob}

In this section, we will set up an interpolation problem which we study in this paper.

\midasi{Notations}  Let $p,q$ be two base variables satisfying constraints $|p|, |q|<1$. 
We denote by $\th{x}$ the Jacobi theta function with base $p$:
\begin{equation}
\th{x}=\prod_{i=0}^{\infty}(1-x p^i)(1-x^{-1}p^{i+1}), \quad
\th{px}=\th{x^{-1}}=-x^{-1}\th{x}.
\end{equation}
The elliptic Gamma function \cite{Rui} and Pochhammer symbol are defined as
\begin{equation}
\Gamma(x;p,q)=\prod_{i,j=0}^{\infty}\frac{(1-x^{-1}p^{i+1}q^{j+1})}{(1-x p^i q^j)}, \quad
\th{x}_s=\dfrac{\Gamma(q^s x;p,q)}{\Gamma(x;p,q)}=\prod_{i=0}^{s-1}\th{q^ix},
\end{equation}
where the last equality holds for $s \in {\Z}_{\geq 0}$.
We shall use the standard convention
\begin{equation}
\begin{array}l
\Gamma(x_1,\cdots,x_\ell;p,q)=\Gamma(x_1;p,q) \cdots \Gamma(x_\ell;p,q), \\
\th{x_1,\cdots,x_\ell}_s=\th{x_1}_s \cdots \th{x_\ell}_s.
\end{array}
\end{equation}

\midasi{Pad\'e problem}
Let $m,n \in {\Z}_{\geq 0}$, and let $a_1,\cdots,a_6$, $k$ be complex parameters with a constraint:
\begin{equation}\label{eq:akcons}
\prod_{i=1}^6 a_i=k^3.
\end{equation}
In this paper we consider the following interpolation problem:
\begin{equation}\label{eq:padeprob}
Y_s=\dfrac{V(q^{-s})}{U(q^{-s})}, \quad (s=0,1,\cdots, N=m+n),
\end{equation}
specified by the following data:

\noindent$\bullet$ The interpolated sequence $Y_s$ is given by
\begin{equation}\label{eq:Ys}
Y_s=Y(q^{-s})=\prod_{i=1}^6\dfrac{\th{a_i}_s}{\th{\sfrac{k}{a_i}}_s},\quad
Y(x)=\prod_{i=1}^6\dfrac{\Gamma(\sfrac{a_i}{x},\sfrac{k}{a_i};p,q)}{\Gamma(\sfrac{k}{a_ix},a_i;p,q)}.
\end{equation}

\noindent$\bullet$
The interpolating functions $U(x), V(x)$ are defined as
\begin{equation}
U(x)=\sum_{i=0}^n u_i \phi_i(x), \quad
V(x)=\sum_{i=0}^m v_i \chi_i(x),
\end{equation}
with basis 
\begin{equation}\label{eq:phichi}
\begin{array}{c}
\phi_i(x)=\dfrac{T_{a_2}^{-i}T_{a_4}^iY(x)}{Y(x)}
=\dfrac{\th{\sfrac{a_4}{x},\sfrac{k}{q^ia_4x}}_i}{\th{\sfrac{a_2}{q^ix},\sfrac{k}{a_2x}}_i}
\dfrac{\th{\sfrac{a_2}{q^i},\sfrac{k}{a_2}}_i}{\th{a_4,\sfrac{k}{q^ia_4}}_i},\\
\chi_i(x)=\dfrac{Y(x)}{T_{a_1}^i T_{a_3}^{-i}Y(x)}
=\dfrac{\th{\sfrac{a_3}{q^ix},\sfrac{k}{a_3x}}_i}{\th{\sfrac{a_1}{x},\sfrac{k}{q^ia_1x}}_i}
\dfrac{\th{a_1,\sfrac{k}{q^ia_1}}_i}{\th{\sfrac{a_3}{q^i},\sfrac{k}{a_3}}_i},
\end{array}
\end{equation}
where $T_a : f(a) \mapsto f(q a)$. 

The coefficients $u_i, v_i$ are determined by eq.(\ref{eq:padeprob}) which is linear homogeneous equations. We normalize them as $u_0=1$.

\midasi{Remark on the choice of the bases $\phi_i(x), \chi_i(x)$}
The problem we are considering is a version of PPZ scheme (interpolation with prescribed poles and zeros) \cite{Zhe}.
Note that
\begin{equation}
\begin{array}l
U(x)=\dfrac{U_{\rm num}(x)}{U_{\rm den}(x)}, \quad U_{\rm den}(x)=\th{\sfrac{a_2}{q^nx},\sfrac{k}{a_2x}}_n,\\
V(x)=\dfrac{V_{\rm num}(x)}{V_{\rm den}(x)}, \quad V_{\rm den}(x)=\th{\sfrac{a_1}{x},\sfrac{k}{q^ma_1x}}_m,
\end{array}
\end{equation}
where $U_{\rm num}(x)$, $U_{\rm den}(x)$ (resp.~$V_{\rm num}(x), V_{\rm den}(x)$) are theta functions of order $2n$
(resp.~$2m$). Furthermore, the functions $x^mU_{\rm num}(x)$, $x^nV_{\rm num}(x)$, $x^mU_{\rm den}(x)$, $x^nV_{\rm den}(x)$ 
(and hence $U(x)$, $V(x)$, $\phi_i(x)$, $\chi_i(x)$ also) are  ``symmetric" : $F(k/qx)=F(x)$.
We will fix the denominator $U_{\rm den}$ (resp.~$V_{\rm den}$) as above in order to specify the prescribed zeros (resp.~poles).
For the numerator $U_{\rm num}$ (resp.~$V_{\rm num}$), contrarily, one may take any basis of theta functions 
as far as they have the same order, same quasi $p$-periodicity, and same symmetry under 
$x \leftrightarrow \sfrac{k}{qx}$ as $U_{\rm den}$ (resp.~$V_{\rm den}$).
In this sense, the choice of the basis $\phi_i$, $\chi_i$ in eq.(\ref{eq:phichi}) is not so essential 
for general argument, however, we will see that it is convenient for explicit expression 
of the functions $U(x)$, $V(x)$ in section \ref{sect:det}.

\midasi{Parameters of the elliptic Painlev\'e equation}
The elliptic Painlev\'e equation is specified by a generic configuration of 8 points on ${\P}^1\times{\P}^1$.
We parametrize them as $\Big(f_*(\xi_i), g_*(\xi_i)\Big)_{i=1,\ldots, 8}$, where
\begin{equation}\label{eq:fgpara}
f_*(x)=\dfrac{\th{\sfrac{c_2}{x},\sfrac{\kappa_1}{c_2 x}}}{\th{\sfrac{c_1}{x},\sfrac{\kappa_1}{c_1 x}}},\quad
g_*(x)=\dfrac{\th{\sfrac{c_4}{x},\sfrac{\kappa_2}{c_4 x}}}{\th{\sfrac{c_3}{x},\sfrac{\kappa_2}{c_3 x}}},
\end{equation}
and $c_i$ are parameters independent of $x$. 
The functions $f_*(x), g_*(x)$ satisfy $f_*(x)=f_*(\sfrac{\kappa_1}{x})$, $g_*(x)=g_*(\sfrac{\kappa_2}{x})$, and they
give a parametrization of an elliptic curve of degree (2,2).\footnote{The choice of parameters 
$c_1,\ldots, c_4$ (and over all normalization of $f_*(x)$, $g_*(x)$) 
is related to the fractional linear transformations on ${\P}^1\times{\P}^1$.}
We define functions $\hhf{f}{x}$ and $\hhg{g}{x}$ as
\begin{equation}\label{eq:FGdef}
\hhf{f}{x}=\th{\frac{c_1}{x},\frac{\kappa_1}{c_1 x}}f-\th{\frac{c_2}{x},\frac{\kappa_1}{c_2 x}},\quad
\hhg{g}{x}=\th{\frac{c_3}{x},\frac{\kappa_2}{c_3 x}}g-\th{\frac{c_4}{x},\frac{\kappa_2}{c_4 x}}.
\end{equation}
Note that $\hhf{f}{x}=0 \Leftrightarrow f=f_*(x)$ and $\hhg{g}{x}=0 \Leftrightarrow g=g_*(x)$.

In this paper, the Painlev\'e equation appears with the following parameters
\begin{equation}\label{eq:8pt}
(\kappa_1,\kappa_2)=(k,\dfrac{k^2}{a_1}), \quad
(\xi_1,\ldots,\xi_8)=(\sfrac{k}{q}, k q^{m+n},\sfrac{k}{a_1q^m},\sfrac{a_2}{q^n},a_3,a_4,a_5,a_6).
\end{equation}
Note that $\kappa_1^2\kappa_2^2=q \xi_1\cdots\xi_8$ due to the constraint (\ref{eq:akcons}). 

\section{Contiguity relations}\label{sect:cont}

Here, we will derive two fundamental contiguity relations\footnote{
Since the contiguity relations (\ref{eq:L2}),(\ref{eq:L3}) are similar to the linear relations of the $R_{\rm II}$ chain \cite{SZ},
it may be possible to derive them as a reduction of three discrete-time non-autonomous Toda chain by
using the method in \cite{Tsuji}.
} satisfied by the functions
$V(x)$, $Y(x)U(x)$.

\midasi{Special direction $T$ of deformation}
For any quantity (or function) $F$ depending on variables $k,a_1,\cdots,a_6,m,n,\cdots$, 
we denote by $\ev{F}=T(F)$ its parameter shift along a special direction $T$:
\begin{equation}
T: (k,a_1,\cdots,a_6,m,n) \mapsto (kq,\sfrac{a_1}{q},a_2,a_3q,\cdots,a_6q,m+1,n-1).
\end{equation}
This special direction is chosen so that 
$T: (\kappa_1, \kappa_2, \xi_i) \mapsto (\kappa_1q, \kappa_2 q^3, \xi_i q)$ and
the corresponding elliptic Painlev\'e equation will take a simple form.

\begin{prop}\label{prop:cont} The functions $y(x)=V(x), Y(x)U(x)$ satisfy the following contiguity relations:
\begin{equation}\label{eq:L2}
L_2: \dfrac{\hhg{g}{\sfrac{kx}{a_1}}\prod_{i=1}^8\th{\sfrac{\xi_i}{x}}}{\th{\sfrac{k}{a_1 x},\sfrac{k}{qx}}}y(x)
-\dfrac{\hhg{g}{x}\prod_{i=1}^8\th{\sfrac{k}{x\xi_i}}}{\th{\sfrac{a_1}{x},\sfrac{q}{x}}}y(\sfrac{x}{q})
-\dfrac{{C_0}\hhf{f}{x}\th{\sfrac{k}{x^2},\sfrac{a_1}{qx},\sfrac{kq}{a_1 x}}}{x}\ev{y}(x)=0,
\end{equation}
\begin{equation}\label{eq:L3}
L_3: \hhg{g}{\sfrac{kqx}{a_1}}\th{\sfrac{k}{qx},\sfrac{kq}{a_1x}}\ev{y}(x)
-\hhg{g}{x}\th{\sfrac{1}{x},\sfrac{a_1}{q^2x}}\ev{y}(qx)
-\dfrac{{C_1}\ev{\hhf{f}{qx}}\th{\sfrac{k}{qx^2}}}{x \th{\sfrac{k}{a_1 x},\sfrac{a_1}{qx}}}y(x)=0,
\end{equation}
where $C_0,C_1,f, g$ are some constants w.r.t. $x$.
\end{prop}

\prf
We put ${\bf y}(x)=\left[\begin{array}{c}V(x)\\ Y(x)U(x)\end{array}\right]$ and define
the Casorati determinants $D_i$ as
\begin{equation}\label{eq:casorati-dets}
\begin{array}l
D_1(x):=\det[{\bf y}(x),{\bf y}(\sfrac{x}{q})],\\
D_2(x):=\det[{\bf y}(qx),{\bf y}(x)],\\
D_3(x):=\det[\ev{{\bf y}}(x),{\bf y}(x)],\\
D_4(x):=\det[\ev{{\bf y}}(x),{\bf y}(\sfrac{x}{q})].
\end{array}
\end{equation}
Then the desired contiguity relations are obtained from the identity
\begin{equation}
\begin{array}l
D_1(x) \ev{y}(x)-D_4(x)y(x)+D_3(x)y(\sfrac{x}{q})=0,\\
D_4(qx) \ev{y}(x)-D_3(x) \ev{y}(qx)-\ev{D_2(x)}y(x)=0,
\end{array}
\end{equation}
by using the formulae for $D_i$ given in the next Lemma \ref{lemma:Dform}.\qed

\begin{lemma}\label{lemma:Dform} The determinants (\ref{eq:casorati-dets}) take the following form:
\begin{equation}\label{eq:D-form}
\begin{array}l
D_1(x)={\mathcal N}(x)Y(x)c\dfrac{\th{\sfrac{k}{x^2},\sfrac{q}{x},\sfrac{a_1}{x}}\hhf{f}{x}}
{x \th{\sfrac{k}{q x},\sfrac{k}{x a_1}}\prod_{i=1}^8\th{\sfrac{k}{x \xi_i}}},\\
D_2(x)={\mathcal N}(x)Y(x)c\dfrac{\th{\sfrac{k}{q^2 x^2},\sfrac{k}{q^2 x},\sfrac{k}{q x a_1}}\hhf{f}{q x}}
{qx \th{\sfrac{1}{x},\sfrac{a_1}{q x}}\prod_{i=1}^8\th{\sfrac{\xi_i}{q x}}},\\
D_3(x)={\mathcal N}(x)Y(x)c'\dfrac{\hhg{g}{x}}{\th{\sfrac{k}{q x},\sfrac{k}{x a_1},\sfrac{k q}{x a_1},\sfrac{a_1}{q x}}},\\
\displaystyle
D_4(x)={\mathcal N}(x)Y(x)c'\dfrac{\th{\sfrac{q}{x},\sfrac{a_1}{x}} \hhg{g}{\sfrac{k x}{a_1}}}
{\th{\sfrac{k}{q x},\sfrac{k}{q x},\sfrac{k}{x a_1},\sfrac{k}{x a_1},\sfrac{k q}{x a_1},\sfrac{a_1}{q x}}}
\prod_{i=1}^8\dfrac{\th{\sfrac{\xi_i}{x}}}{\th{\sfrac{k}{x \xi_i}}},
\end{array}
\end{equation}
where
\begin{equation}
{\mathcal N}(x)=\dfrac{\th{\sfrac{1}{q^{m+n}x},\sfrac{k}{qx}}_{m+n+1}}{U_{\rm den}(x) V_{\rm den}(x)}.
\end{equation}
\end{lemma}

\prf
The functions $U(x)$, $V(x)$ and, due to the constraint (\ref{eq:akcons}), the function $Y(x)$
are elliptic ($p$-periodic) functions in $x$.
Hence the ratios $\frac{D_i(x)}{Y(x)}$ are also elliptic.
They are of order $2m+2n+$(small corrections) and have sequences of zeros and poles represented as
$\th{\sfrac{1}{q^{m+n}x},\sfrac{k}{qx}}_{m+n+1}$ and $U_{\rm den} V_{\rm den}$ modulo corrections at the boundaries of the sequence. 
Then we can compute the ratios $\frac{D_i(x)}{Y(x)}$, and each of them are determined up to $2$ unknown constants.
In the computation, the following relations are useful (they are derived by a straightforward computation)
\begin{equation}
G(x):=\dfrac{Y(qx)}{Y(x)}=\prod_{i=1}^6 \dfrac{\th{\sfrac{k}{a_iq x}}}{\th{\sfrac{a_i}{qx}}},
\end{equation}
\begin{equation}
K(x):=\dfrac{\ev{Y}(x)}{Y(x)}=\displaystyle
\dfrac{\th{\sfrac{k}{a_1},\sfrac{k}{a_2},\sfrac{a_1}{q},\sfrac{kq}{a_1}}}
{\th{\sfrac{k}{a_1x},\sfrac{k}{a_2x},\sfrac{a_1}{qx},\sfrac{kq}{a_1x}}}
\prod_{i=3}^{6} \dfrac{\th{\sfrac{a_i}{x}}}{\th{a_i}},
\end{equation}
\begin{equation}
{\mathcal N}(\sfrac{k}{qx})=\dfrac{q x^2}{k}{\mathcal N}(x),
\end{equation}
and
\begin{equation}
\dfrac{{{\mathcal N}(qx)}}{{\mathcal N}(x)}=
\dfrac{\th{\sfrac{q}{x},\sfrac{q^N k}{x},\sfrac{a_1}{x},\sfrac{k}{q^m a_1 x},\sfrac{k}{a_2 x},\sfrac{a_2}{q^n x}}}
{\th{\sfrac{1}{q^{N+1}x},\sfrac{k}{qx},\sfrac{q^m a_1}{x},\sfrac{k}{a_1 x},\sfrac{q^n k}{a_2 x},\sfrac{a_2}{x}}}.
\end{equation}

\noindent
$\bullet$ {\it Computation of $D_1(x)$, $D_2(x)$}:
First, we count the degree of the elliptic function
\begin{equation}
\dfrac{D_1(x)}{Y(x)}=\dfrac{1}{G(\sfrac{x}{q})}V(x)U(\sfrac{x}{q})-V(\sfrac{x}{q})U(x).
\end{equation}
Substituting 
\begin{equation}
\begin{array}l
U(\sfrac{x}{q})=\dfrac{U_{\rm num}(\sfrac{x}{q})}{U_{\rm den}(\sfrac{x}{q})}
=\dfrac{\th{\sfrac{k}{a_2 x},\sfrac{a_2}{q^n x}}}{\th{\sfrac{q^n k}{a_2 x},\sfrac{a_2}{x}}}\dfrac{U_{\rm num}(\sfrac{x}{q})}{U_{\rm den}(x)},\\
V(\sfrac{x}{q})=\dfrac{V_{\rm num}(\sfrac{x}{q})}{V_{\rm den}(\sfrac{x}{q})}
=\dfrac{\th{\sfrac{k}{q^ma_2 x},\sfrac{a_1}{x}}}{\th{\sfrac{k}{a_1 x},\sfrac{q^m a_1}{x}}}\dfrac{V_{\rm num}(\sfrac{x}{q})}{V_{\rm den}(x)},
\end{array}
\end{equation}
we have
\begin{equation}
\begin{array}l
\dfrac{D_1(x)}{Y(x)}=\dfrac{1}{U_{\rm den}(x)V_{\rm den}(x)}\dfrac{\th{\sfrac{a_1}{x}}}{\th{\sfrac{k}{a_1 x}}} \\
\displaystyle \phantom{\dfrac{D_1(x)}{Y(x)}=}\times \Big\{
\dfrac{\th{\sfrac{a_2}{q^n x}}}{\th{\sfrac{q^n k}{a_2 x}}}\prod_{i=3}^6 
\dfrac{\th{\sfrac{a_i}{x}}}{\th{\sfrac{k}{a_i x}}}V_{\rm num}(x)U_{\rm num}(\sfrac{x}{q})
-\dfrac{\th{\sfrac{k}{q^m a_1 x}}}{\th{\sfrac{q^m a_1}{x}}}U_{\rm num}(x)V_{\rm num}(\sfrac{x}{q})
\Big\}.
\end{array}
\end{equation}
The function $D_1(x)/Y(x)$ is $p$-periodic function of order $2m+2n+6$ with denominator
\begin{equation}
U_{\rm den}(x) \Big\{V_{\rm den}(x)\dfrac{\th{\sfrac{k}{a_1 x}}}{\th{\sfrac{a_1}{x}}}\Big\}
\th{\sfrac{q^m a_1}{x}}\th{\sfrac{q^n k}{a_2 x}}\prod_{i=3}^6 \th{\sfrac{k}{a_i x}}.
\end{equation}
Next, we study the zeros. 
When $x$ and $\sfrac{x}{q}$ are both in the Pad\'e interpolation grid (i.e. for $x=1,q^{-1},\ldots,q^{-N+1}$), 
it follows obviously that $D_1(x)=0$.
Noting the symmetry properties
\begin{equation}
U(\sfrac{k}{qx})=U(x), \quad V(\sfrac{k}{qx})=V(x), \quad G(\sfrac{k}{qx})=\dfrac{1}{G(\sfrac{x}{q})}, 
\end{equation}
we have
\begin{equation}
\dfrac{D_1(\sfrac{k}{x})}{Y(\sfrac{k}{x})}=G(\sfrac{x}{q})U(x)V(\sfrac{x}{q})-U(\sfrac{x}{q})V(x)=
-G(\sfrac{x}{q})\dfrac{D_1(x)}{Y(x)}.
\end{equation}
Then it follows that $D_1(x)=0$ at $x=k,kq,\cdots,kq^{N-1}$ and furthermore, due to the relation
${\bf y}(x)={\bf y}(\sfrac{x}{q})$ for $x^2=k$, we have $D_1(x)=0$ at $x^2=k$ (i.e. $x=\pm \sqrt{k},\pm \sqrt{kp}$). 
As a result, the function $X(x)$ defined by
\begin{equation}
D_1(x)={\mathcal N}(x)Y(x)\dfrac{\th{\sfrac{k}{x^2},\sfrac{q}{x},\sfrac{a_1}{x}}}{x \th{\sfrac{k}{q x},\sfrac{k}{x a_1}}\prod_{i=1}^8\th{\sfrac{k}{x \xi_i}}}X(x)
\end{equation}
is a theta function of degree $2$ such that $X(\sfrac{x}{p})=X(\sfrac{k}{x})=\sfrac{x^2}{k}X(x)$, hence it can be written as
$X(x)=c F_f(x)$ by suitable constants $c,f$. 
$D_2$ is easily obtained since $D_2(x)=D_1(qx)$.

\noindent
$\bullet$ {\it Computation of $D_3(x)$, $D_4(x)$}:
First we note a relation between $D_3(x)$ and $D_4(x)$.
Using $U(\sfrac{k}{qx})=U(x)$, $\ev{U}(\sfrac{k}{x})=\ev{U}(x)$ and similar relations for $V(x)$ we have
\begin{equation}\label{eq:D3D4}
\begin{array}l
\dfrac{D_3(\sfrac{k}{qx})}{Y(\sfrac{k}{qx})}=U(\sfrac{k}{qx})\ev{V}(\sfrac{k}{qx})-K(\sfrac{k}{qx})\ev{U}(\sfrac{k}{qx})V(\sfrac{k}{qx})\\
\quad=U(x)\ev{V}({qx})-K(\sfrac{k}{qx})\ev{U}({qx})V(x)\\
\quad=\dfrac{G(x)}{Y(qx)}\Big\{Y(x)U(x)\ev{V}({qx})-\dfrac{K(\sfrac{k}{qx})}{G(x)}Y(qx)\ev{U}({qx})V(x)\Big\}\\
\quad=G(x)\dfrac{D_4(qx)}{Y(qx)},
\end{array}
\end{equation}
where we have used the relation $\dfrac{K(\sfrac{k}{qx})}{G(x)}=K(qx)$ at the last step.

Let us compute $D_3(x)$. Substituting the relation
\begin{equation}
\begin{array}l
\ev{U}(x)=\dfrac{\ev{U_{\rm num}}(x)}{\ev{U_{\rm den}}(x)}=\th{\sfrac{k}{a_2 x},\sfrac{a_2}{q^n x}}\dfrac{\ev{U_{\rm num}}(x)}{U_{\rm den}(x)},\\
\ev{V}(x)=\dfrac{\ev{V_{\rm num}}(x)}{\ev{V_{\rm den}}(x)}
=\dfrac{\th{\sfrac{k}{q^m a_1 x}}}{\th{\sfrac{a_1}{q x},\sfrac{k}{a_1 x},\sfrac{q k}{a_1 x}}}\dfrac{\ev{V_{\rm num}}(x)}{V_{\rm den}(x)},
\end{array}
\end{equation}
into 
\begin{equation}
\dfrac{D_3(x)}{Y(x)}=U(x)\ev{V}(x)-K(x) \ev{U}(x)V(x),
\end{equation}
we have
\begin{equation}
\begin{array}l
\dfrac{D_3(x)}{Y(x)}=\dfrac{1}{U_{\rm den}(x)V_{\rm den}(x)} \dfrac{1}{\th{\sfrac{k}{a_1 x},\sfrac{q k}{a_1 x},\sfrac{a_1}{q x}}}\\
\displaystyle \phantom{\dfrac{D_4(x)}{Y(x)}=}\times \Big\{\th{\sfrac{a_2}{q^n x},\sfrac{a_3}{x},\cdots,\sfrac{a_6}{x}} V_{\rm num}(x)\ev{U_{\rm num}}(x)-
\th{\sfrac{k}{q^m a_1 x}}\ev{V_{\rm num}}(x)U_{\rm num}(x)\Big\},
\end{array}
\end{equation}
Hence, $\dfrac{D_3(x)}{Y(x)}$ is of degree $2m+2n+3$.

$D_3(x)$ has zeros at $x=1,q^{-1},\ldots,q^{-N}$ and $x=k,qk,\ldots,q^{N-1}k$, where the latter zeros follow from those of $D_4(x)$ through eq.(\ref{eq:D3D4}).
Hence, we obtain
\begin{equation}
D_3(x)={\mathcal N}(x)Y(x)\dfrac{1}{\th{\sfrac{k}{q x},\sfrac{k}{x a_1},\sfrac{k q}{x a_1},\sfrac{a_1}{q x}}}Z(x),
\end{equation}
where $Z(x)$ is a theta function of degree 2 such as $Z(\sfrac{x}{p})=Z(\sfrac{k^2}{a_1 x})=\sfrac{a_1 x^2}{k^2}Z(x)$, namely
$Z(x)=c' G_g(x)$ for some $c'$ and $g$ as desired. $D_4(x)$ is derived by the relation (\ref{eq:D3D4}). \qed

\begin{coro} For any pair $i,j \in \{3,4,5,6\}$ we have
\begin{equation}\label{eq:fgbyUV}
\begin{array}l
\dfrac{\alpha(a_i)}{\alpha(a_j)}\dfrac{\hhf{f}{a_i}}{\hhf{f}{a_j}}=\dfrac{U(a_i)V(a_i/q)}{U(a_j)V(a_j/q)},\quad
\dfrac{\beta(a_i)}{\beta(a_j)}\dfrac{\hhg{g}{a_i}}{\hhg{g}{a_j}}=\dfrac{U(a_i)\ev{V}(a_i)}{U(a_j)\ev{V}(a_j)},
\end{array}
\end{equation}
where
\begin{equation}
\begin{array}l
\alpha(x)={\mathcal N}(x)
\dfrac{\th{\sfrac{k}{x^2},\sfrac{q}{x},\sfrac{a_1}{x}}}{x \th{\sfrac{k}{q x},\sfrac{k}{x a_1}}\prod_{i=1}^8\th{\sfrac{k}{x \xi_i}}},\quad
\beta(x)={\mathcal N}(x)
\dfrac{1}{\th{\sfrac{k}{q x},\sfrac{k}{x a_1},\sfrac{k q}{x a_1},\sfrac{a_1}{q x}}}.
\end{array}
\end{equation}
\end{coro}

\prf
By the definition of $D_1, D_3$, we have for $x=a_i$ ($i=3,4,5,6$)
\begin{equation}
\begin{array}l
\dfrac{D_1(x)}{Y(x)}=\dfrac{1}{G({x}/{q})}V(x)U(\sfrac{x}{q})-U(x)V(\sfrac{x}{q})=-U(x)V(\sfrac{x}{q}),\\
\dfrac{D_3(x)}{Y(x)}=\ev{V}(x)U(x)-K(x)\ev{U}(x)V(x)=U(x)\ev{V}(x).
\end{array}
\end{equation}
Then, from the first and the third equation of (\ref{eq:D-form}), one has eq.(\ref{eq:fgbyUV}).\qed

The formulae (\ref{eq:fgbyUV}) are convenient in order to obtain $f, g$ from $U(x)$, $V(x)$.

\section{Elliptic Painlev\'e equation}\label{sect:painl}

In this section, we study the eqs.(\ref{eq:L2}),(\ref{eq:L3}) for generic variables
$f,g$ apart from the Pad\'e problem, and 
prove that the variables $f,g$ satisfy the elliptic Painlev\'e equation.

\begin{thm} If the eqs.(\ref{eq:L2}),(\ref{eq:L3}) are compatible, then the variables $f,g$ and $\ev{f},\ev{g}$ should be related by
\begin{equation}\label{eq:fev}
\dfrac{\hhf{f}{x} \ev{\hhf{f}{q x}}}{\hhf{f}{\sfrac{x a_1}{k}}\ev{\hhf{f}{\sfrac{q^2 x a_1}{k}}}}
=\prod_{i=1}^8\dfrac{\th{\sfrac{\xi_i}{x}}}{\th{\sfrac{k^2}{x \xi_i a_1}}}, \quad {\rm for} \quad  g=g_*(x),
\end{equation}
and
\begin{equation}\label{eq:gev}
\dfrac{\hhg{g}{x} \ev{\hhg{g}{q x}}}{\hhg{g}{\sfrac{k q x}{a_1}}\ev{\hhg{g}{\sfrac{kqx}{a_1}}}}
=\prod_{i=1}^8\dfrac{\th{\sfrac{\xi_i}{x}}}{\th{\sfrac{k}{qx \xi_i}} },
\quad {\rm for} \quad \ev{f}=\ev{f_*}(qx).
\end{equation}
\end{thm}

\prf
From equations $\ev{L_2}|_{x \to qx}$ (\ref{eq:L2}) and $L_3$ (\ref{eq:L3}) we have
\begin{equation}
\begin{array}l
\dfrac{\ev{\hhg{g}{\sfrac{kqx}{a_1}}}\prod_{i=1}^8\th{\sfrac{\xi_i}{x}}}{\th{\sfrac{kq}{a_1x},\sfrac{k}{qx}}}\ev{y}(qx)
=\dfrac{\ev{\hhg{g}{qx}}\prod_{i=1}^8\th{\sfrac{k}{qx\xi_i}}}{\th{\sfrac{a_1}{q^2x},\sfrac{1}{x}}}\ev{y}(x),\\
\hhg{g}{\sfrac{kqx}{a_1}}\th{\sfrac{k}{qx},\sfrac{kq}{a_1x}}\ev{y}(x)
=\hhg{g}{x}\th{\sfrac{1}{x},\sfrac{a_1}{q^2x}}\ev{y}(qx),
\end{array}
\end{equation}
for $\ev{f}=\ev{f_*}(qx)$, hence we have eq.(\ref{eq:gev}).

For $g=g_*(x)$, we have from eqs.(\ref{eq:L2}), (\ref{eq:L3}) that
\begin{equation}
\begin{array}l
\dfrac{\hhg{g}{\sfrac{kx}{a_1}}\prod_{i=1}^8\th{\sfrac{\xi_i}{x}}}{\th{\sfrac{k}{a_1x},\sfrac{k}{qx}}}y(x)
=\dfrac{{C_0}\hhf{f}{x}\th{\sfrac{k}{x^2},\sfrac{a_1}{qx},\sfrac{kq}{a_1x}}}{x}\ev{y}(x),\\
\hhg{g}{\sfrac{kqx}{a_1}}\th{\sfrac{k}{qx},\sfrac{kq}{a_1x}}\ev{y}(x)
=\dfrac{{C_1}\ev{\hhf{f}{qx}}\th{\sfrac{k}{qx^2}}}{x \th{\sfrac{k}{a_1x},\sfrac{a_1}{qx}}}y(x),
\end{array}
\end{equation}
hence
\begin{equation}\label{eq:C0C1}
\hhg{g}{\sfrac{kx}{a_1}}\hhg{g}{\sfrac{kqx}{a_1}}\prod_{i=1}^8\th{\sfrac{\xi_i}{x}}
=\dfrac{w}{x^2}\hhf{f}{x}\ev{\hhf{f}{qx}}\th{\sfrac{k}{x^2},\sfrac{k}{qx^2}},
\end{equation}
where $w=C_0C_1$.
The eq.(\ref{eq:C0C1}) holds also by replacing $x \to \sfrac{k^2}{a_1 x}$ since $g_*(x)=g_*(\sfrac{k^2}{a_1 x})$, 
Taking a ratio eq.(\ref{eq:C0C1}) with eq.(\ref{eq:C0C1})$|_{x \to \sfrac{k^2}{a_1 x}}$ we have eq.(\ref{eq:fev}).\qed

The next Lemma \ref{lem:fggeom} shows that the relations (\ref{eq:fev}),(\ref{eq:gev}) are equivalent to the
time evolution equation for the elliptic Painlev\'e.\footnote{
Since the elliptic Painlev\'e equation \cite{Sakai} is rather complicated, its concise expressions have been pursued
by several authors (e.g. \cite{MSY},\cite{Murata},\cite{ORG}). The system (\ref{eq:fev}),(\ref{eq:gev}) is supposed to be the simplest one.}
\begin{lemma}\label{lem:fggeom}
The solution $\ev{f}$ of eq.(\ref{eq:fev}) is a rational function of $(f,g)$ of degree $(1,4)$, which is characterized by
the following conditions:
(i) its numerator and denominator have $8$ zeros at $f=f_*(\xi), g=g_*(\xi)$, 
(ii) if $f=f_*(u)$, $g=g_*(u)$ $(u\neq \xi)$ then $\ev{f}=\ev{f_*}(\sfrac{a_1 u}{k})$.
Similarly, by eq.(\ref{eq:gev}), $\ev{g}$ is uniquely given as a rational function of $(\ev{f},g)$ of degree $(4,1)$, satisfying the conditions
(i') it has $8$ points of indeterminacy at $\ev{f}=\ev{f_*}(q\xi), g=g_*(\xi)$, 
(ii') if $\ev{f}=\ev{f_*}(qu)$, $g=g_*(u)$ $(u\neq \xi)$ then $\ev{g}=\ev{g_*}(\sfrac{q^3 k u}{a_1})=\ev{g_*(\sfrac{q k u}{a_1})}$.
\end{lemma}

\prf
Written in the form
\begin{equation}
\hhf{f}{x} \ev{\hhf{f}{q x}} \prod_{i=1}^8 \th{\sfrac{k^2}{x \xi_i a_1}}
=\hhf{f}{\sfrac{x a_1}{k}} \ev{\hhf{f}{\sfrac{q^2 x a_1}{k}}} \prod_{i=1}^8\th{\sfrac{\xi_i}{x}},
\end{equation}
the eq.(\ref{eq:fev}) is quasi $p$-periodic in $x$ of degree (apparently) $12$ with symmetry under 
$x \leftrightarrow \sfrac{k^2}{a_1 x}$. Since it is divisible by a factor $\th{\sfrac{k^2}{a_1x^2}}$, 
it is effectively of degree $8$. Then the solution $\ev{f}$ of this equation takes the form
\begin{equation}
\ev{f}=\dfrac{A(x) f+B(x)}{C(x)f+D(x)},
\end{equation}
where the coefficients $A(x), \ldots, D(x)$ are $x \leftrightarrow \sfrac{k^2}{a_1 x}$-symmetric 
$p$-periodic functions of degree $8$, namely polynomials of $g=g_*(x)$ of degree $4$.
Hence $\ev{f}$ is a rational function of $(f,g)$ of degree $(1,4)$. The conditions (i), (ii) 
are obvious by the form of eq.(\ref{eq:fev}).
The structure of the solution $\ev{g}=\ev{g}(\ev{f},g)$ of the eq.(\ref{eq:gev}) is similar.\qed

\midasi{Remark on the geometric characterization of the solutions $f,g$}
As a consequence of the above results, the variables $f,g$ obtained from the Pad\'e problem give
special solutions of the elliptic Painlev\'e equation.
Since they are (B\"acklund transformations of) the terminating hypergeometric solution \cite{KMNOY1} \cite{KMNOY2}, 
they have the following geometric characterization.
Let $C_1$ be a curve of degree $(2n,2n+1)$ passing through the 8 points
$\Big(f_*(\xi_i),g_*(\xi_i)\Big)_{i=1}^8$ in eqs.(\ref{eq:fgpara}),(\ref{eq:8pt}) 
with multiplicity $n(1^8)+(0,1,1,0,0,0,0,0)$.
Similarly, Let $C_2$ be a curve of degree $(2m+2,2m+1)$ passing through the 8 points
with multiplicity $m(1^8)+(0,1,0,1,1,1,1,1)$. $C_1$ and $C_2$ are unique rational
curves. Except for the assigned 8 points, there exist unique unassigned intersection 
point $(f,g) \in C_1 \cap C_2$ which is the solution.

\section{Lax formalism}\label{sect:lax}

In this section, we prove that the elliptic Painlev\'e equation (\ref{eq:fev}),(\ref{eq:gev}) 
are sufficient for the compatibility of eqs.(\ref{eq:L2}),(\ref{eq:L3}).

\begin{figure}[h]
\begin{center}
\setlength{\unitlength}{1.2mm}
\begin{picture}(40,35)(10,5)
\put(8,10){$y(\sfrac{x}{q})$}
\put(29,10){$y(x)$}
\put(50,10){$y(qx)$}
\put(8,30){$\ev{y}(\sfrac{x}{q})$}
\put(29,30){$\ev{y}(x)$}
\put(50,30){$\ev{y}(qx)$}
\put(30,4){$L_1$}
\put(20,19){$L_2$}
\put(40,19){$L_3$}
\put(30,37){$L_1'$}
\put(14,8){\line(1,0){35}}
\put(14,13){\line(1,0){15}}
\put(29,13){\line(0,1){15}}
\put(14,34){\line(1,0){35}}
\put(34,28){\line(1,0){15}}
\put(34,13){\line(0,1){15}}
\end{picture}
\caption{Lax equations}
\label{fig:lax}
\end{center}
\end{figure}
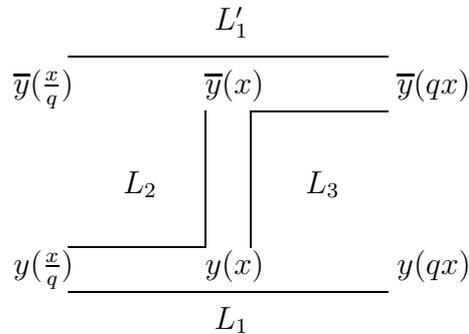

Solving $\ev{y}(x)$ and $\ev{y}(qx)$ from eqs.$L_2$, $L_2|_{x \to qx}$ and plugging them into $L_3$, one has the
following difference equation (Fig.\ref{fig:lax}):
\begin{equation}\label{eq:L1}
\begin{array}l
L_1 : \dfrac{\th{\sfrac{k}{a_1x},\sfrac{k}{qx}}\prod_{i=1}^8\th{\sfrac{k}{x\xi_i}}}{\hhf{f}{x}\th{\sfrac{k}{x^2},\sfrac{a_1}{x},\sfrac{q}{x}}}y(\sfrac{x}{q})
+\dfrac{q \th{\sfrac{1}{x},\sfrac{a_1}{qx}}\prod_{i=1}^8\th{\sfrac{\xi_i}{qx}}}{\hhf{f}{qx}\th{\sfrac{k}{q^2x^2},\sfrac{k}{q^2x},\sfrac{k}{a_1qx}}}y(qx)\\
+\Bigg\{\dfrac{{w}\ev{\hhf{f}{qx}}\th{\sfrac{k}{qx^2}}}{x^2\hhg{g}{x}\hhg{g}{\sfrac{kqx}{a_1}}}
-\dfrac{q\hhg{g}{qx}\prod_{i=1}^8\th{\sfrac{k}{qx\xi_i}}}{\hhf{f}{qx}\hhg{g}{\sfrac{kqx}{a_1}}\th{\sfrac{k}{q^2x^2}}}
-\dfrac{\hhg{g}{\sfrac{kx}{a_1}}\prod_{i=1}^8\th{\sfrac{\xi_i}{x}}}{\hhf{f}{x}\hhg{g}{x}\th{\sfrac{k}{x^2}}}\Bigg\}y(x)=0.
\end{array}
\end{equation}
The pairs of equations $\{L_1,L_2\}$, $\{L_1,L_3\}$ and $\{L_2,L_3\}$ are equivalent with each other.

The above expression $L_1$ (\ref{eq:L1}) contains variables $f,g,\ev{f}, w$. We will rewrite and
characterize it in terms of $f, g$ only. This characterization is a key of the proof of the compatibility.
To do this, we first note the following
\begin{lemma}
The factor $w$ satisfying the relation (\ref{eq:C0C1}) is explicitly given by $(f,g)$ as 
\begin{equation}\label{eq:wsol}
w=C \dfrac{\ev{f}_{\rm den}(f,g)}{\varphi(f,g)},
\end{equation}
where $\ev{f}_{\rm den}(f,g)$ is a polynomial of degree $(1,4)$ defined as the denominator of
the rational function $\ev{f}=\ev{f}(f,g)$, and $\varphi(f,g)$ is the defining
polynomial of the degree $(2,2)$ curve parametrized by $f_*(x), g_*(x)$, and
$C$ is a constant independent of $f,g,x$. 
\end{lemma}

\prf
The relation (\ref{eq:C0C1}) follows from eq.(\ref{eq:wsol}) by using
\begin{equation}
\varphi\big{|}_{g=g_*(x)}=C'\dfrac{\hhf{f}{x}\hhf{f}{\sfrac{a_1 x}{k}}}{{g_*}_{\rm den}(x)^2},
\end{equation}
\begin{equation}
\Big(\ev{f}_{\rm den} \ev{f_*}_{\rm num}(qx)-\ev{f}_{\rm num} \ev{f_*}_{\rm den}(qx)\Big)\Big{|}_{g=g_*(x)}=C''
\dfrac{\hhf{f}{\sfrac{a_1x}{k}}\prod_{i=1}^8\th{\sfrac{\xi_i}{x}}}{{g_*}_{\rm den}(x)^4},
\end{equation}
where $C'$, $C''$ are constants, ${g_*}_{\rm den}(x)=\th{\sfrac{c_3}{x},\sfrac{k^2}{a_1c_3 x}}$ is the denominator of $g_*(x)$, 
and similarly 
$\ev{f_*}_{\rm den}(x)=\th{\sfrac{c_{1}}{x},\sfrac{kq}{c_{1} x}}$,
$\ev{f_*}_{\rm num}(x)=\th{\sfrac{c_{2}}{x},\sfrac{kq}{c_{2} x}}$.
\qed

\begin{lemma}\label{lem:L1geom}
In terms of variables $f,g$, the eq.(\ref{eq:L1}) is represented as a
polynomial equation $L_1(f,g)=0$ of degree $(3,2)$ characterized \footnote{This geometric characterization of the difference equation $L_1$ is essentially the same as that
in \cite{Ysigma}.} by the following vanishing conditions at:
(1) 10 points $(f_*(u),g_*(u))$ where $u=\xi$,
$qx$ and $\sfrac{k}{x}$, 
(2) $2$ more points $(f,g)$ such as 
\begin{equation}\label{eq:Qx}
f=f_*(x), \quad \dfrac{y(x)}{y(\sfrac{x}{q})}\dfrac{\hhg{g}{\sfrac{kx}{a_1}}}{\hhg{g}{x}}=
\frac{\th{\sfrac{k}{a_1x},\sfrac{k}{qx}}}{\th{\sfrac{a_1}{x},\sfrac{q}{x}}}\prod_{i=1}^8 \dfrac{\th{\sfrac{k}{\xi_i x}}}{\th{\sfrac{\xi_i}{x}}},
\end{equation}
and
\begin{equation}\label{eq:Qqx}
f=f_*(qx), \quad \dfrac{y(qx)}{y(x)}\dfrac{\hhg{g}{\sfrac{kqx}{a_1}}}{\hhg{g}{qx}}=
\frac{\th{\sfrac{k}{a_1qx},\sfrac{k}{q^2x}}}{\th{\sfrac{a_1}{qx},\sfrac{1}{x}}}\prod_{i=1}^8 \dfrac{\th{\sfrac{k}{\xi_i qx}}}{\th{\sfrac{\xi_i}{qx}}}.
\end{equation}
\end{lemma}

\prf
Due to the eq.(\ref{eq:C0C1}), 
the residue of $L_1$ at the apparent pole $g=g_*(x)$ vanishes.
Replacing $x$ with $\sfrac{k}{qx}$ in eq.(\ref{eq:C0C1}) and using the relations 
$\hhf{f}{\sfrac{k}{x}}=\frac{x^2}{k}\hhf{f}{x}$ and
$\hhg{g}{\sfrac{k^2}{a_1x}}=\frac{a_1x^2}{k^2}\hhg{g}{x}$, we have
\begin{equation}
q x^2 \hhg{g}{x}\hhg{g}{qx}\prod_{i=1}^8\th{\sfrac{k}{q\xi_i x}}
=w\hhf{f}{qx}\ev{\hhf{f}{qx}}\th{\sfrac{k}{qx^2},\sfrac{k}{q^2x^2}},
\end{equation}
hence, the residue of $L_1$ at $g=g_*(\sfrac{kqx}{a_1})=g_*(\sfrac{k}{qx})$ also vanishes.
From these vanishing of residues and the eq.(\ref{eq:wsol}), the L.H.S of eq.(\ref{eq:L1}) turns out to be a polynomial in $(f,g)$
of degree $(3,2)$, after multiplying by $\hhf{f}{x}\hhf{f}{qx}\varphi$.
Check of the vanishing conditions (1),(2) are easy.\qed

In a similar way, solving $y(\sfrac{x}{q}), y(x)$ form $L_3$, $L_3|_{x \to x/q}$ and substituting them into $L_2$, one has
\begin{equation}\label{eq:L1p}
\begin{array}l
L_1': \dfrac{\th{\sfrac{1}{x},\sfrac{a_1}{q^2 x}} \prod_{i=1}^8\th{\sfrac{\xi_i}{x}}}{\th{\sfrac{k}{q x^2},\sfrac{k}{q x},\sfrac{k q}{x a_1}}\ev{\hhf{f}{qx}}}\ev{y}(q x) 
+\dfrac{\th{\sfrac{k}{x},\sfrac{k q^2}{x a_1}} \prod_{i=1}^8 \th{\sfrac{k}{x \xi_i}}}{q \th{\sfrac{k q}{x^2},\sfrac{q}{x},\sfrac{a_1}{q x}}\ev{\hhf{f}{x}}}\ev{y}(\sfrac{x}{q})\\
+\Big\{
\dfrac{w \th{\sfrac{k}{x^2}}\hhf{f}{x}}{x^2 \hhg{g}{x}\hhg{g}{\sfrac{k x}{a_1}}}
-\dfrac{\hhg{g}{\sfrac{x}{q}} \prod_{i=1}^8\th{\sfrac{k}{x \xi_i}} }{q \th{\sfrac{k q}{x^2}}\ev{\hhf{f}{x}}\hhg{g}{\sfrac{kx}{a_1}}}
-\dfrac{\hhg{g}{\sfrac{k q x}{a_1}}\prod_{i=1}^8\th{\sfrac{\xi_i}{x}}}{\th{\sfrac{k}{q x^2}}\ev{\hhf{f}{qx}}\hhg{g}{x}}\Big\}\ev{y}(x) =0.
\end{array}
\end{equation}
By the similar analysis as $L_1$, we have the following
\begin{lemma}\label{lem:L1'geom}
In terms of variables $\ev{f},g$, the eq.(\ref{eq:L1p}) is represented as a
polynomial equation $L_1'(\ev{f},g)=0$ of degree $(3,2)$ characterized by the following vanishing conditions at:
(1) 10 points $(\ev{f_*}(qu),g_*(u))$ where $u=\xi$,
$\sfrac{x}{q}$ and $\sfrac{k}{qx}$. 
(2) $2$ more points $(\ev{f},g)$ such as 
\begin{equation}\label{eq:Qx'}
\ev{f}=\ev{f_*(x)}, \quad \dfrac{\ev{y}(x)}{\ev{y}(\sfrac{x}{q})}\dfrac{\hhg{g}{\sfrac{x}{q}}}{\hhg{g}{\sfrac{kx}{a_1}}}=
\frac{\th{\sfrac{kq^2}{a_1x},\sfrac{k}{x}}}{\th{\sfrac{a_1}{qx},\sfrac{q}{x}}},
\end{equation}
and
\begin{equation}\label{eq:Qqx'}
\ev{f}=\ev{f_*(qx)}, \quad \dfrac{\ev{y}(qx)}{\ev{y}(x)}\dfrac{\hhg{g}{{x}}}{\hhg{g}{\sfrac{kqx}{a_1}}}=
\frac{\th{\sfrac{kq}{a_1x},\sfrac{k}{qx}}}{\th{\sfrac{a_1}{q^x},\sfrac{1}{x}}}.
\end{equation}
\end{lemma}

\prf
In terms of $(\ev{f},g)$, the gauge factor $w$ (\ref{eq:wsol}) is written as
\begin{equation}\label{eq:wsol1}
w=C''' \dfrac{f_{\rm den}(\ev{f},g)}{\ev{\varphi}(\ev{f},g)},
\end{equation}
where $f_{\rm den}(\ev{f},g)$ is the denominator of
the rational function $f=f(\ev{f},g)$, and $\ev{\varphi}(\ev{f},g)$ is the defining
polynomial of the curve parametrized by $\ev{f}_*(qx), g_*(x)$, and
$C'''$ is a constant.
Then the proof of the Lemma is the same as the proof of the Lemma \ref{lem:L1geom}.\qed

\begin{prop}
The eq.(\ref{eq:L1p}) expressed in terms of $(\ev{f}, \ev{g})$ is equivalent with the transformation $T(L_1)=\ev{L_1}$ of eq.(\ref{eq:L1}).
\end{prop}

\prf
This fact is a consequence of Lemmas \ref{lem:fggeom}, \ref{lem:L1geom} and \ref{lem:L1'geom}.
The geometric proof in the $q$-difference case \cite{Yimrn} is also available here (see Lemmas 4.2 - 4.6 in \cite{Yimrn}).\qed

\section{Determinant formulae}\label{sect:det}

In this section, we present explicit determinant formulae for the solutions $U(x)$, $V(x)$
of the interpolation problem (\ref{eq:padeprob}).

\begin{thm}\label{thm:UVdet} Interpolating rational functions $U(x)$, $V(x)$ have the following
determinant expressions:
\begin{equation}\label{eq:UVdet}
\begin{array}l
U(x)={\rm const.}\left| \begin{array}{ccc}
m^U_{0,0}&\cdots&m^U_{0,n}\\
\vdots&\ddots&\vdots\\
m^U_{n-1,0}&\cdots&m^U_{n-1,n}\\
\phi_0(x)&\cdots&\phi_n(x)
\end{array}\right|,\quad
V(x)={\rm const.}\left| \begin{array}{ccc}
m^V_{0,0}&\cdots&m^V_{0,m}\\
\vdots&\ddots&\vdots\\
m^V_{m-1,0}&\cdots&m^V_{m-1,m}\\
\chi_0(x)&\cdots&\chi_m(x)
\end{array}\right|,
\end{array}
\end{equation}
where
\begin{equation}
\begin{array}l
m^U_{ij}={}_{12}V_{11}(q^{-1}k, q^{-N}, q^{N- i - 1}a_1, q^{-j}a_2, q^ia_3, q^ja_4, a_5, a_6;q),\\
m^V_{ij}={}_{12}V_{11}(q^{-1}k, q^{-N}, q^{-j}\sfrac{k}{a_1}, q^{N-i-1}\sfrac{k}{a_2}, q^j\sfrac{k}{a_3}, q^i\sfrac{k}{a_4}, \sfrac{k}{a_5}, \sfrac{k}{a_6};q),
\end{array}
\end{equation}
and ${}_{n+5}V_{n+4}$ (${}_{n+3}E_{n+2}$ in convention of \cite{KMNOY1}) is the very-well poised, balanced elliptic hypergeometric series \cite{DJKMO}\cite{Spi}:
\begin{equation}
{}_{n+5}V_{n+4}(u_0;u_1,\cdots,u_n;z)=\sum_{s=0}^{\infty}\dfrac{\th{u_0 q^{2s}}}{\th{u_0}}\prod_{j=0}^{n} \dfrac{\th{u_j}_s}{\th{qu_0/u_j}_s}z^s.
\end{equation}
\end{thm}

\prf
In general, the solution of interpolation problem
\begin{equation}
V(x_s)=Y_sU(x_s), \quad s=0,\cdots,N.
\end{equation}
is written by the following determinants:
\begin{equation}\label{eq:generalUdet}
U(x)=\left|\begin{array}{cccccc}
\chi_0(x_0)&\cdots&\chi_m(x_0)&Y_0\phi_0(x_0)&\cdots&Y_0\phi_n(x_0)\\
\vdots&\ddots&\vdots&\vdots&\ddots&\vdots\\
\chi_0(x_N)&\cdots&\chi_m(x_N)&Y_N\phi_0(x_N)&\cdots&Y_N\phi_n(x_N)\\
0&\cdots&0&\phi_0(x)&\cdots&\phi_n(x)
\end{array}\right|,
\end{equation}
and
\begin{equation}
V(x)=\left|\begin{array}{cccccc}
\chi_0(x_0)&\cdots&\chi_m(x_0)&Y_0\phi_0(x_0)&\cdots&Y_0\phi_n(x_0)\\
\vdots&\ddots&\vdots&\vdots&\ddots&\vdots\\
\chi_0(x_N)&\cdots&\chi_m(x_N)&Y_N\phi_0(x_N)&\cdots&Y_N\phi_n(x_N)\\
\chi_0(x)&\cdots&\chi_m(x)&0&\cdots&0
\end{array}\right|.
\end{equation}
We apply these formulae for $Y_s$, $\phi_i(x)$, $\chi_i(x)$ given by (\ref{eq:Ys}), (\ref{eq:phichi}) 
and $x_s=q^{-s}$.
Note that $\phi_i(x_s)$, $\chi_i(x_s)$ can be written as
\begin{equation}
\begin{array}l
\phi_i(x_s)=\dfrac{\th{\sfrac{k}{a_2},\sfrac{k}{a_4},q^{-i}a_2,q^ia_4}_s}{\th{a_2,a_4,q^i\sfrac{k}{a_2},q^{-i}\sfrac{k}{a_4}}_s},\\
\chi_i(x_s)=\dfrac{\th{a_1,a_3,q^{-i}\sfrac{k}{a_1},q^i\sfrac{k}{a_3}}_s}{\th{\sfrac{k}{a_1},\sfrac{k}{a_3},q^ia_1,q^{-i}a_3}_s}.
\end{array}
\end{equation}
To rewrite the determinant in eq.(\ref{eq:generalUdet}), we use the multiplication by a matrix
\begin{equation}
L=\left[\begin{array}{c|c}
\big(L_{ij}\big)_{i,j=0}^N&\\ \hline
&1
\end{array}
\right]
\end{equation}
from the left, where
\begin{equation}
L_{ij}=\dfrac{\th{q^{2j-1}k}}{\th{q^{-1}k}}
\dfrac{\th{q^{-1}k,q^{-N}, q^{N-i-1}a_1, \sfrac{k}{a_1}, q^ia_3,\sfrac{k}{a_3}}_j}
{\th{q, q^Nk,q^{-N+i+1}\sfrac{k}{a_1},a_1,q^{-i}\sfrac{k}{a_3},a_3}_j} q^j.
\end{equation}
For the last $n+1$ columns, we have
\begin{equation}
\sum_{s=0}^N L_{is}Y_s\phi_j(x_s)
={}_{12}V_{11}(q^{-1}k;q^{-N},q^{N-i-1}a_1,q^{-j}a_2,q^ia_3,q^j a_4,a_5,a_6;q)=m^U_{ij}.
\end{equation}
For the first $m+1$ columns, we have
\begin{equation}\label{eq:Lchi}
\sum_{s=0}^N L_{is}\chi_j(x_s)
={}_{10}V_9(q^{-1}k;q^{-N},q^{N-i-1}a_1,q^{-j}\sfrac{k}{a_1},q^ia_3,q^j\sfrac{k}{a_3};q).
\end{equation}
Using the Frenkel-Turaev summation formula ($u_1\cdots u_5=q u_0^2$, $u_5=q^{-n}$) \cite{FT}\cite{Spi}:
\begin{equation}
{}_{10}V_{9}(u_0;u_1,\cdots,u_5;q)=
\dfrac{\th{qu_0,\sfrac{qu_0}{u_1u_2},\sfrac{qu_0}{u_1u_3},\sfrac{qu_0}{u_2u_3}}_n}
{\th{\sfrac{qu_0}{u_1},\sfrac{qu_0}{u_2},\sfrac{qu_0}{u_3},\sfrac{qu_0}{u_1u_2u_3}}_n},
\end{equation}
the expression (\ref{eq:Lchi}) can be evaluated  as
\begin{equation}
\dfrac{\th{k,q^{-N+i+j+1},q^{-N+1}\sfrac{k}{a_1a_3},q^{j-i}\sfrac{a_1}{a_3}}_N}
{\th{q^{-N+j+1}\sfrac{1}{a_3},q^{-i}\sfrac{k}{a_3},q^ja_1,q^{-N+i+1}\sfrac{k}{a_1}}_N},
\end{equation}
and it vanishes for $0\leq i+j <N$. Hence, we obtain the formula for $U(x)$ in (\ref{eq:UVdet})
by Laplace expansion. The case for function $V(x)$ is similar.\qed

Theorem \ref{thm:UVdet} supplies also formulae for special solutions $f,g$ of the elliptic Painlev\'e equation through eq.(\ref{eq:fgbyUV}). 
Moreover we have
\begin{lemma}
For $i,j \in \{3,4,5,6\}$, the ratios in eq.(\ref{eq:fgbyUV}) have following simple form 
\begin{equation}\label{eq:taushift}
\dfrac{U(a_i)}{U(a_j)}=\dfrac{c_i T_{a_2}^{-1}T_{a_i}(\tau^U)}{c_j T_{a_2}^{-1}T_{a_j}(\tau^U)}, \quad
\dfrac{V(a_i/q)}{V(a_j/q)}=\dfrac{c_i' T_{a_1}T_{a_i}^{-1}(\tau^V)}{c_j' T_{a_1}T_{a_j}^{-1}(\tau^V)}, 
\end{equation}
where $\tau^U=\det (m^U_{i,j})_{i,j=0}^{n-1}$, $\tau^V=\det (m^V_{i,j})_{i,j=0}^{m-1}$,
\[
\begin{array}l
c_3=q^{\sfrac{n(n-1)}{2}}\dfrac{(q^{-n}\sfrac{k}{a_3},q)_n(a_3,q)_n(q^{-m-n+1}\sfrac{a_3}{a_1},q)_n(q^{m+1}\sfrac{a_1a_3}{k},q)_n}{(\sfrac{k}{a_2a_3},q)_n(\sfrac{qa_3}{a_2},q)_n(q^{-m-n+1}\sfrac{a_3}{a_1},q^2)_n(q^{m+n}\sfrac{a_1a_3}{k},1)_n},\\
c_4=\dfrac{(q^{-n}\sfrac{k}{a_4},q)_n(a_4,q)_n}{(\sfrac{k}{a_2a_4},1)_n(\sfrac{qa_4}{a_2},q^2)_n},\qquad 
c_i=\dfrac{(\sfrac{k}{qa_i},1)_n(a_i,1)_n}{(\sfrac{k}{a_2a_i},q)_n(\sfrac{qa_i}{a_2},q)_n},\quad (i=5,6),
\end{array}
\]
$(x,v)_n=\prod_{i=0}^{n-1} \vartheta_p(x v^i)$
and $(c_3',c_4',c_5',c_6')=(c_4,c_3,c_5,c_6){\Big |}_{(m,n,a_1,\ldots,a_6)\mapsto (n,m,\sfrac{k}{a_2},\sfrac{k}{a_1},\sfrac{k}{a_4},\sfrac{k}{a_3},\sfrac{k}{a_5},\sfrac{k}{a_6})}$.
\end{lemma}
\prf
Since $\phi_i(a_4)=\delta_{i,0}$ $(i\geq 0)$, we have 
\begin{equation}
\dfrac{U(a_4)}{\rm const.}=\det (m^U_{i,j+1})_{i,j=0}^{n-1}=T_{a_2}^{-1}T_{a_4}(\tau^U).
\end{equation}
Using the symmetry of $U(x)$ in parameters $a_3,\ldots, a_6$, the first relation of eq.(\ref{eq:taushift})
follows. The second relation is similar.\qed

The determinant expressions for the special solutions have been known for various (discrete) Painlev\'e equations
(see \cite{Masuda} \cite{Rains} for example). Our method using Pad\'e interpolation gives a simple and direct way to obtain them.

\appendix
\section{Affine Weyl group actions}

Here we give a derivation of the Painlev\'e equation (\ref{eq:fev}), (\ref{eq:gev}) 
from the affine Weyl group actions.\cite{MSY} \cite{Yimrn}

Define multiplicative transformations $s_{ij}$, $c$, $\mu_{ij}$, $\nu_{ij}$ ($1\leq i\neq j \leq 8$)
acting on variables  $h_1, h_2, u_1, \ldots, u_8$ as
\begin{equation}\label{eq:pic-action}
\begin{array}l
s_{ij} = \{u_i \leftrightarrow u_j\},\qquad
c= \{h_1 \leftrightarrow h_2\},\\
\mu_{ij} = \{h_1 \mapsto \frac{h_1h_2}{u_iu_j}, \quad
u_i \mapsto \frac{h_2}{u_j}, \quad u_j \mapsto \frac{h_2}{u_i}\},\\
\nu_{ij} = \{h_2 \mapsto \frac{h_1h_2}{u_iu_j}, \quad
u_i \mapsto \frac{h_1}{u_j}, \quad u_j \mapsto \frac{h_1}{u_i}\}.
\end{array}
\end{equation}
These actions generate the affine Weyl group of type $E^{(1)}_8$ with the following simple reflections:
\begin{equation}
\begin{array}{cccccccccccccccccc}
&&&&s_{12}\\
&&&&\vert\\
c&-&\mu_{12}&-&s_{23}&-&s_{34}&-&\cdots&-&s_{78} &&.
\end{array}
\end{equation}
We extend the actions bi-rationally on variables $(f,g)$.
The nontrivial actions are as follows:
\begin{equation}
c(f)=g, \quad c(g)=f, \quad
\mu_{ij}(f)=\tilde{f},\quad
\nu_{ij}(g)=\tilde{g},
\end{equation}
where, $\tilde{f}=\tilde{f}_{ij}$ and $\tilde{g}=\tilde{g}_{ij}$ are rational functions in $(f,g)$ defined by
\begin{equation}
\frac{\tilde{f}-\mu_{ij}(f_i)}{\tilde{f}-\mu_{ij}(f_j)}=\frac{(f-f_i)(g-g_j)}{(f-f_j)(g-g_i)},\quad
\frac{\tilde{g}-\nu_{ij}(g_i)}{\tilde{g}-\nu_{ij}(g_j)}=\frac{(g-g_i)(f-f_j)}{(g-g_j)(f-f_i)},
\end{equation}
$(f_i,g_i)=(f_\star(u_i), g_\star(u_i))$, and
\begin{equation}
f_\star(z)=\dfrac{\th{\sfrac{d_2}{z},\sfrac{h_1}{d_2 z}}}{\th{\sfrac{d_1}{z},\sfrac{h_1}{d_1 z}}},\quad
g_\star(z)=\dfrac{\th{\sfrac{d_2}{z},\sfrac{h_2}{d_2 z}}}{\th{\sfrac{d_1}{z},\sfrac{h_2}{d_1 z}}},
\end{equation}
as in eq.(\ref{eq:fgpara}).
As a rational function of $(f,g)$, $\tilde{f}$ is characterized by the following properties: 
(i) it is of degree $(1,1)$ with indeterminate points
$(f_i,g_i)$, $(f_j,g_j)$, (ii) it maps generic points on the elliptic curve $(f_\star(z),g_\star(z))$ to 
$\dfrac{\th{\sfrac{d_2}{z},\sfrac{h_1h_2}{d_2 z u_1 u_2}}}{\th{\sfrac{d_1}{z},\sfrac{h_1h_2}{d_1 z u_i u_j}}}$. 
Using this geometric characterization, we have
\begin{equation}\label{eq:mu}
\mu_{ij}\Big\{\dfrac{{\mathcal F}_f(\sfrac{h_1 z}{h_2})}{{\mathcal F}_f(z)}\Big\}=\dfrac{\th{\sfrac{u_i}{z},\sfrac{u_j}{z}}}{\th{\sfrac{h_2}{u_i z},\sfrac{h_2}{u_j z}}}
\dfrac{{\mathcal F}_f(\sfrac{h_1 z}{h_2})}{{\mathcal F}_f(z)}, \quad {\rm for} \quad g=g_\star(z),
\end{equation}
where the functions ${\mathcal F}_f(z)$ (and ${\mathcal G}_g(z)$) are defined in a similar way as eq.(\ref{eq:FGdef})
\begin{equation}
{\mathcal F}_{f}(z)=\th{\frac{d_1}{z},\frac{h_1}{d_1 z}}f-\th{\frac{d_2}{z},\frac{h_1}{d_2 z}},\quad
{\mathcal G}_{g}(z)=\th{\frac{d_1}{z},\frac{h_2}{d_1 z}}g-\th{\frac{d_2}{z},\frac{h_2}{d_2 z}}.
\end{equation}

Let us consider the following compositions \cite{MSY}
\begin{equation}
r=s_{12}\mu_{12}s_{34}\mu_{34}s_{56}\mu_{56}s_{78}\mu_{78}, \quad
T=rcrc.
\end{equation}
Their actions on variables $(h_i, u_i)$ are given by
\begin{equation}
\begin{array}{lllll}
r(h_1)=v h_2, &
r(h_2)=h_2, &
r(u_i)=\frac{h_2}{u_i},\\
T(h_1)=q h_1 v^2, &
T(h_2)=q^{-1} h_2 v^2, &
T(u_i)=u_i v,\\
\end{array}
\end{equation}
where $v=qh_2/h_1$, $q=h_1^2h_2^2/(u_1\cdots u_8)$.
{}From eq.(\ref{eq:mu}) and
$r(\sfrac{h_1}{h_2})=\sfrac{q h_2}{h_1}$, the evolution $T(f)=rcrc(f)=r(f)$ is determined  as
\begin{equation}
\dfrac{{\mathcal F}_f(z)}{{\mathcal F}_f(\sfrac{h_1 z}{h_2})}\dfrac{T({\mathcal F}_f)(\sfrac{q h_2 z}{h_1})}{T({\mathcal F}_f)(z)}
=\prod_{i=1}^8 \dfrac{\th{\sfrac{u_i}{z}}}{\th{\sfrac{h_2}{u_i z}}}, \quad {\rm for} \quad g=g_\star(z).
\end{equation}
Similarly, since $cTc=T^{-1}$, $T^{-1}(g)$ is determined by
\begin{equation}
\dfrac{{\mathcal G}_g(z)}{{\mathcal G}_g(\sfrac{h_2 z}{h_1})}\dfrac{T^{-1}({\mathcal G}_g)(\sfrac{q h_1 z}{h_2})}{T^{-1}({\mathcal G}_g)(z)}
=\prod_{i=1}^8 \dfrac{\th{\sfrac{u_i}{z}}}{\th{\sfrac{h_1}{u_i z}}}, \quad {\rm for} \quad f=f_\star(z).
\end{equation}
By a re-scaling of variables $(h_i,u_i,d_i)=(\kappa_i \lambda^2, \xi_i \lambda, c_i\lambda)$
with $\lambda=(h_1^3 h_2^{-1})^\frac{1}{4}$, we have ${\mathcal F}_f(z)=F_f(\sfrac{z}{\lambda})$, $T({\mathcal F}_f)(z)=T(F_f)(\sfrac{\kappa_1}{\kappa_2}\sfrac{z}{\lambda})$ and so on, since $T(\lambda)=\sfrac{h_2}{h_1}\lambda$.
Then the above equations take the form (\ref{eq:fev}), (\ref{eq:gev}), by putting $z=\lambda x$.

\vskip10mm

\noindent
{\bf Acknowledgment.} 
This work was partially supported by JSPS Grant-in-aid for Scientific Research (KAKENHI) 21340036, 22540224 and 19104002.


\vskip5mm

\begin{thebibliography}{A}
\bibitem{DJKMO}
E.~Date, M.~Jimbo, A.~Kuniba, T.~Miwa, M.~Okado, 
{\it Exactly solvable SOS models II: Proof of the star-triangle 
relation and combinatorial identities}, 
Adv. Stud. Pure Math. {\bf 16}, Academic Press, Boston, (1988) 17--122.
%
\bibitem{FT}
I.~B.~Frenkel, V.~G.~Turaev, 
{\it Elliptic solutions of the Yang-Baxter equation and modular hypergeometric functions}, 
in: I.~Arnold, et al. (Eds.), The Arnold-Gelfand Mathematical Seminars, Birkh\"auser, Boston, (1997) 171--204. 
%
\bibitem{JS}
M.~Jimbo and H.~Sakai, 
{\it A $q$-analog of the sixth Painlev\'e equation}, 
Lett. Math. Phys. {\bf 38} (1996) 145-154.
%
\bibitem{KMNOY1}
K.~Kajiwara, T.~Masuda, M.~Noumi, Y.~Ohta and Y.~Yamada,
{\it ${}_{10}E_9$ solution to the elliptic Painlev\'e equation},
J. Phys. {\bf A36} (2003) L263-L272.
%
\bibitem{KMNOY2}
K.~Kajiwara, T.~Masuda, M.~Noumi, Y.~Ohta and Y.~Yamada,
{\it Hypergeometric solutions to the $q$-Painlev\'e equations}, 
IMRN 2004 {\bf 47} (2004) 2497-2521.
%
\bibitem{Magnus}
A.~Magnus, 
{\it Painlev\'e-type differential equations for the recurrence coefficients of semi-classical orthogonal polynomials}, 
J. Comput. Appl. Math. {\bf 57} (1995) 215-237.
%
\bibitem{Masuda}
T.~Masuda,
{\it Hypergeometric $\tau$-functions of the q-Painlev\'e system of type $E^{(1)}_8$},
Ramanujan J. {\bf 24} (2011) 1--31.
%
\bibitem{MSY}
M.~Murata, H.~Sakai, and J.~Yoneda,
{\it Riccati solutions of discrete Painlev\'e equations with Weyl group symmetry of type $E_8^{(1)}$},
J. Math. Phys. {\bf 44} (2003) 1396--1414.
%
\bibitem{Murata}
M.~Murata,
{\it New expressions for discrete Painlev\'e equations}, 
Funkcial. Ekvac. {\bf 47} (2004) 291--305. 
%
\bibitem{ORG}
Y.~Ohta, A.~Ramani and B.~Grammaticos, 
{\it An affine Weyl group approach to the eight-parameter discrete Painlev\'e equation},
J. Phys. {\bf A34} (2001) 10523-10532.
%
\bibitem{QRT}
G.~R.~W.~Quispel, J.~A.~G.~Roberts, C.~J.~Thompson,
{\it Integrable mappings and soliton equations},
Phys. Lett. {\bf A126} (1988) 419--421.
%
\bibitem{Rui}
S.~N.~M.~Ruijsenaars, 
{\it First order analytic difference equations and integrable quantum systems},
J. Math. Phys. {\bf 38} (1997) 1069--1146. 
%
\bibitem{Rains}
E.~Rains, 
{\it Recurrences for elliptic hypergeometric integrals},
Rokko Lectures in Mathematics {\bf 18} (2005) 183--199.
%
\bibitem{Sakai}
H.~Sakai, 
{\it Rational surfaces with affine root systems and geometry of
the Painlev\'e equations},
Commun. Math. Phys. {\bf 220} (2001) 165--221.
%
\bibitem{Spi}
V.~P.~Spiridonov,
{\it Essays on the theory of elliptic hypergeometric functions}, 
Uspekhi Matematicheskikh Nauk {\bf 63} (2008) 3--72.
{\it Classical elliptic hypergeometric functions and their applications},
Rokko Lectures in Mathematics {\bf 18} (2005) 253--287.
%
\bibitem{SZ}
V.~Spiridonov and A.~Zhedanov,
{\it Spectral transformation chains and some new biorthogonal rational functions}, 
Commun. Math. Phys. {\bf 210} (2000) 49--83.
%
\bibitem{Tsuji}
S.~Tsujimoto,
{\it Determinant solutions of the nonautonomous discrete Toda equation associated with the deautonomized discrete KP hierarchy},
J. Syst. Sci. Complex. {\bf 23} (2010) 153--176.
%
\bibitem{Yfe}
Y.~Yamada,
{\it Pad\'e method to Painlev\'e equations},
Funkcial. Ekvac. {\bf 52} (2009) 83--92.
%
\bibitem{Ysigma} 
Y.~Yamada, 
{\it A Lax formalism for the elliptic difference Painlev\'e equation},
SIGMA {\bf 5} (2009) 042 (15pp).
%
\bibitem{Yimrn}
Y.~Yamada,
{\it Lax formalism for $q$-Painlev\'e equations with affine Weyl group symmetry of type $E^{(1)}_n$},
IMRN 2011 {\bf 17} (2011) 3823--3838.
%
\bibitem{Zhe}
A.~S.~Zhedanov,
{\it Pad\'e interpolation table and biorthogonal rational functions},
Rokko Lectures in Mathematics {\bf 18} (2005) 323--363.
%
\end{thebibliography}
\end{document}